\documentclass[aps,pre]{revtex4}
\usepackage{graphicx}
\usepackage{amsfonts}
\usepackage{amssymb}
\usepackage{amsmath}

\newcommand{\beq}{\begin{equation}}
\newcommand{\eeq}{\end{equation}}

\def\1{\'{a}}
\def\3{\'{e}}
\def\9{\'{\i}}
\def\0{\'{o}}
\def\8{\'{u}}

\def\be{\begin{equation}}
\def\ee{\end{equation}}
\def\ba{\begin{eqnarray}}
\def\ea{\end{eqnarray}}
\def\bas{\begin{eqnarray*}}
\def\eas{\end{eqnarray*}}

\begin{document}

\title{Dynamical Barrier for the Formation of Solitary Waves in Discrete Lattices}

\author{P.G. Kevrekidis and J.A. Espinola-Rocha}
\affiliation{Department of Mathematics and Statistics, University of
Massachusetts, Amherst, MA 01003}
\author{Y. Drossinos}
\affiliation{European Commission, Joint Research Centre, I-21020
Ispra (Va), Italy \\ and \\School of Mechanical and Systems
Engineering, University of Newcastle upon Tyne, Newcastle upon Tyne
NE1 7RU, United Kingdom}
\author{A. Stefanov}
\affiliation{Department of Mathematics, University of Kansas, 1460
Jayhawk Blvd., Lawrence, KS 66045--7523}

\date{\today}

%
%
\begin{abstract}
We consider the problem of the existence of a dynamical barrier of
``mass'' that needs to be excited on a lattice site to lead to the
formation and subsequent persistence of localized modes for a
nonlinear Schr{\"o}dinger lattice. We contrast the existence of a
dynamical barrier with its absence in the static theory of localized
modes in one spatial dimension. We suggest an energetic criterion
that provides a sufficient, but not necessary, condition on the
amplitude of a single-site initial condition required to form a
solitary wave. We show that this effect is not one-dimensional by
considering its two-dimensional analog.
The existence of a sufficient condition for the excitation of
localized modes in the non-integrable, discrete, nonlinear
Schr{\"o}dinger equation is
compared to the dynamics of excitations in the integrable, both
discrete and continuum, version of the nonlinear Schr{\"o}dinger equation.
\end{abstract}

\maketitle

\section{Introduction}

In the past few years, there has been an explosion of interest in
discrete models that has been summarized in numerous recent reviews
\cite{reviews}. This growth has been spurted by numerous
applications of dynamical lattice nonlinear models in areas as
diverse as the nonlinear optics of waveguide arrays \cite{optics},
the dynamics of Bose-Einstein condensates in periodic potentials
\cite{bec_reviews}, micro-mechanical models of cantilever arrays
\cite{sievers}, or even simple models of the complex dynamics of the
DNA double strand \cite{peyrard}. Perhaps the most prototypical
model among those that emerge in these settings is the, so-called,
discrete nonlinear Schr{\"o}dinger equation (DNLS) \cite{dnls}. DNLS
may arise as a direct model, as a tight binding approximation, or
even as an envelope wave expansion: it is, arguably, one of the most
ubiquitous models in the nonlinear physics of dispersive, discrete
systems.

In at least one of these settings (namely, in the nonlinear optics
of waveguide arrays with the focusing nonlinearity), the feature
that will be of interest to the present work has been observed
experimentally. In particular, it has been noted, to the best of our
knowledge firstly in Ref.~\cite{mora}, that when an injected beam of
light into one waveguide had low intensity, then the beam dispersed
through quasi-linear propagation. On the other hand, in the same
work, experiments with high intensity of the input beam led to the
first example of formation of discrete solitary waves in waveguide
arrays. A very similar ``crossover'' from linear to nonlinear
behavior was also observed very recently in arrays of waveguides
with the defocusing nonlinearity~\cite{rosberg}. The common feature
of both works is that they used the DNLS equation as the supporting
model to illustrate this behavior at a theoretical/numerical level.
However, this crossover phenomenon is certainly not purely discrete
in nature. Perhaps the most famous example of a nonlinear wave
equation that possesses such a threshold is the {\it integrable}
continuum nonlinear Schr{\"o}dinger equation \cite{sulem}.
Specifically, it is well-known that, e.g., in the case of a square
barrier of initial conditions of amplitude $V_0$ and width $L$, the
product $V_0 L$ determines the nature of the resulting soliton, and
if it is sufficiently small the initial condition disperses without
the formation of a solitonic structure \cite{ZS}.
On the other hand, the existence of the threshold is not a purely
one-dimensional feature either. For instance, experiments on the
formation of solitary waves in two-dimensional photorefractive
crystals show that low intensities lead to diffraction, whereas higher
intensities induce localization~\cite{neshev,fleischer}. Moreover,
similar phenomena were observed even in the formation of
higher-order excited structures such as vortices (as can be inferred
by carefully inspecting the results of
Refs.~\cite{neshev2,fleischer2}). It should be mentioned that the
latter field of light propagation in photorefractive crystals is
another major direction of current research in nonlinear optics;
see, e.g., Ref.~\cite{moti} for a recent review.

This crossover behavior between linear and nonlinear dynamics may be
understood qualitatively rather simply. In the case of power law
nonlinearities of order $p$, which are relevant in these settings, a
small intensity O$(\epsilon)$, where $\epsilon \ll 1$, yields a
nonlinear contribution O$(\epsilon^p)$ that is negligible with
respect to the linear terms of the equation. On the other hand, if
$\epsilon \gg 1$, the opposite will be true and the nonlinear terms
will dominate the linear ones, yielding essentially nonlinear
behavior. A key question regards the details of this crossover and
what determines its more precise location for an appropriately
parametrized initial condition. This is the question we address
herein. We argue that the problem related to the experiments
described above reduces, at the mathematical level, to a DNLS
equation with a Kronecker-$\delta$ initial condition parametrized by
its amplitude. Then, a well defined value of the
initial-state amplitude exists such that initial states with higher
amplitude always give rise to localized modes. The condition
may be determined by comparing the energy of the initial state with the
energy of the localized excitations that the model supports.
This sufficient, but not necessary, condition for the formation of
localized solitary waves provides an intuitively and physically appealing
interpretation of the dynamics that is in very good agreement with our
numerical observations.
We also consider variants of this process
in different settings: for reasons of completeness, we present
it also in the continuum NLS equation, noting the significant
differences that the latter case has from the present one.
As yet another example of very different (from both its non-integrable
sibling and its continuum limit) dynamical behavior, we also present
the case of the integrable discrete NLS (so-called Ablowitz-Ladik
\cite{al1,al2,trub}) model.
In addition to the one-dimensional DNLS lattice, we also
consider the two-dimensional case where the
role of both energy and beam power (mathematically the squared
$l^2$-norm) become apparent. We should note here that our tool of choice for visualizing
the ``relaxational process'' (albeit in a Hamiltonian system)
of the initial condition will be energy-power diagrams.
Such diagrams have proven very helpful in visualizing the dynamics
of initial conditions in a diverse host of nonlinear wave equations.
In particular, they have been used in the nonlinear homogeneous
systems such as birefringent media and nonlinear couplers as is
discussed in Chapters 7 and 8 of Ref. \cite{akm1}. They have also been used
in a form closely related to the present work (but in the continuum
case; see also the discussion below)
for general nonlinearities in dispersive wave equations in \cite{akm2},
while they have been used to examine the migration of localized
excitations in DNLS equations in \cite{rumpf}.

Our presentation is structured as follows. In section II,
we present the analytically tractable theory of the integrable ``relatives''
of the present model: we review the known theory for the NLS model
and develop its analog for the integrable discrete NLS case.
Then, in section III, we
present our analytical and numerical results
in the one- and two-dimensional DNLS equation. In the last section, we
summarize our findings and present our conclusions, as well as highlight
some important questions for future studies.

\section{Threshold Conditions for the Integrable NLS Models}

\subsection{The Continuum NLS Model}

For reasons of completeness of the presentation and to
compare and contrast the results of the non-integrable case that
is at the focus of the present work, we start by summarizing the
threshold conditions for the continuum NLS model \cite{ZS}.
For the focusing NLS equation
\begin{equation}
iu_t = - \frac{1}{2}u_{xx}  - |u|^2u ,
\label{ceq1}
\end{equation}
with squared barrier initial data
\begin{equation}
u(x,0)  =   \left\{\!\begin{array}{cc}  V_0, &   -L \leq x \leq  L   \\  0, &  \mbox{otherwise}\end{array}\!\right.,
\label{ceq2}
\end{equation}
(the inverse of) the transmission coefficient, $S_{11}(E)$, which is the first entry of the scattering matrix, is given by
\begin{eqnarray}
\label{trans_coeff}
S_{11}(E) =   \nu(E)\cos(2\nu(E)L) - iE\sin(2\nu(E)L),
\label{ceq3}
\end{eqnarray}
with
$
\nu(E) = \sqrt{E^2 + V_0^2}
$
where $E$ is the spectral parameter and $V_0$ the amplitude of the barrier.
It is well-known that the number of zeros of this coefficient
represents the number of solitons produced by the square barrier initial
condition \cite{ZS}.

It can be proved that the roots of this equation are purely imaginary.
(This initial condition satisfies the single-lobe conditions of Klaus-Shaw potentials,
from which it follows that  the eigenvalues are purely imaginary \cite{klaus2}). Let us define $\eta \geq 0$ and use
$
E=   i \eta.
$
Then, Eq. (\ref{trans_coeff}) becomes
\begin{eqnarray} \label{trasc_eq}
\sqrt{1-\eta^2}\cos\Big(2V_0\sqrt{1-\eta^2}L\Big) +  \eta\sin\Big(2V_0\sqrt{1-\eta^2}L\Big)  = 0.
\end{eqnarray}
We can verify that Eq. (\ref{trasc_eq}) does not have any roots
(i.e., leads to no solitons in Eq. (\ref{ceq1})) if
\begin{eqnarray}
V_0 < \frac{\pi}{2}.
\label{ceq5}
\end{eqnarray}
Furthermore, the condition to generate $n$ {solitons}, i.e., so that Eq. (\ref{trasc_eq}) has $n$ roots is
\begin{eqnarray}
(2n -1)\frac{\pi}{2}   <   2V_0L  <  (2n +1)\frac{\pi}{2} ,
\label{ceq6}
\end{eqnarray}
or, equivalently, the count of eigenvalues is given by
\begin{eqnarray}
\frac{2}{\pi}V_0 L - \frac{1}{2}    <   n   <  \frac{2}{\pi}V_0 L + \frac{1}{2}.\label{ceq7}
\end{eqnarray}

The limit $V_0 \rightarrow \infty$ together with $ L \rightarrow 0$ can be
reached if we impose
$
 2V_0L  = {\rm const}.
$
In this instance, the number of eigenvalues stays the same.

\subsection{The Ablowitz-Ladik Model}

We now turn to the integrable discretization of Eq. (\ref{ceq1})
and examine its dynamics. The one-dimensional
integrable, discrete nonlinear Schr{\"o}dinger model (so-called
Ablowitz-Ladik (AL) model \cite{al1,al2,trub}) reads:
\begin{eqnarray}
i \dot{u}_n=-\frac{1}{2} (u_{n+1}+u_{n-1}-2 u_n) - \frac{1}{2} |u_n|^2 \left(
u_{n+1} + u_{n-1} \right).
\label{deq10}
\end{eqnarray}

In this case, there exists a  {Lax pair} of linear operators
\cite{al1,al2,trub}
\ba
{\mathcal L}_n  & = &   Z + M_n   \label{discrete_lax_L} , \\
{\mathcal B}_n    & = &  \left(  \frac{z - z^{-1}}{2}\right)^2 D + \frac{1}{2}\left({ZM_n -Z^{-1}M_{n-1}}\right)  - \frac{1}{2}DM_nM_{n-1},\label{discrete_lax_B}
\ea
with the definitions for the matrices
\beq
Z =  \left(\!\begin{array}{cc}  z  & 0 \\ 0 & z^{-1}   \end{array}\!\right),
\hspace{0.5cm}
D  =  \left(\!\begin{array}{cc}  -1 & 0 \\ 0 & 1   \end{array}\!\right),
\hspace{0.5cm}
\mbox{and}
\hspace{0.5cm}
 M_n    =  \left(\!\begin{array}{ccc}  0 & U_n   \\ -U_n^{\star} & 0   \end{array}\!\right),
\eeq
where $z$ is the spectral parameter, and
$U_n = U_n(t)$ is a solution of the equation.   These two
operators (\ref{discrete_lax_L}) and (\ref{discrete_lax_B}) define the
system of differential-difference equations
\ba
\Psi_{n+1} & = &  {\mathcal L}_n\Psi_n ,          \label{AL_eival}\\
i\frac{d}{d\tau}\Psi_n  & = &  {\mathcal B}_n\Psi_n ,   \label{discrete_t_flow}
\ea
for a complex matrix function $\Psi_n$. Then, the compatibility condition of
Eqs. (\ref{AL_eival}) and (\ref{discrete_t_flow})
\[
i\frac{d}{d\tau}\Psi_{n+1}  = i\left.\left(\frac{d}{d\tau}\Psi_m\right)\right|_{m = n+1}
\]
(i.e., the Lax equation) becomes the AL model.
$U_n = U_n(t)$ is referred to
as the potential of the  AL eigenvalue problem.

For $U_n$ decaying rapidly at $\pm\infty$, and for $n  \rightarrow   \pm   \infty$,
from Eq. (\ref{AL_eival}) we have:
\[
\Psi_{n+1} \sim  Z \Psi_{n}.
\]
We normalize this type of  solutions as follows: let $\Psi_{n} $ denote the
solution of Eq. (\ref{AL_eival}) such that
\[
\Psi_{n} \sim  Z^n \hspace{1cm}   \mbox{as}     \hspace{1cm}    n \rightarrow +\infty,
\]
and let $\Phi_{n}$ be the solution of Eq. (\ref{AL_eival}) such that
\[
\Phi_{n}   \sim    Z^n    \hspace{1cm}     \mbox{as}      \hspace{1cm}     n \rightarrow -\infty.
\]
$\Psi_{n} $ and $\Phi_{n} $ are known as the Jost functions.
Each of these
forms a  system of linearly independent solutions of the AL
eigenvalue problem (\ref{AL_eival}). These sets of solutions  are
inter-related by  the {scattering matrix} $S(z)$,
\beq
\Phi_n = \Psi_nS(z). \label{def_scat_matriz}
\eeq
The first column of this equation is given by
\beq
(\Phi_1)_n = S_{11}(z) (\Psi_1)_n + S_{21}(z) (\Psi_2)_n, \label{ecuacion_estrella}
\eeq
where $(\Phi_1)_n$ denotes the first column of $\Phi_n$.
Similar definitions apply to  $(\Psi_1)_n$ and $(\Psi_2)_n$.
Since $\Phi_{n} \sim  Z^n $ as  $   n \rightarrow -\infty$,  then
\[
(\Phi_1)_n \sim z^n \left(\!\begin{array}{c}  1  \\ 0    \end{array}\!\right).
\]
To obtain decay, $(\Phi_1)_n \rightarrow 0$ when $n \rightarrow -\infty$, we
demand
\beq
|z| > 1.
\eeq
On the other hand,
\[
\Psi_n   =   ((\Psi_1)_n ,  (\Psi_2)_n )\sim   \left(\!\begin{array}{cc}  z^n  & 0 \\ 0 & z^{-n}   \end{array}\!\right)  \hspace{1cm}     \mbox{as}      \hspace{1cm}     n \rightarrow \infty.
\]
Therefore, if $|z|>1$ then $(\Psi_1)_n \rightarrow \infty$ and $(\Psi_2)_n \rightarrow   0$, as $n \rightarrow \infty$. Now, from  Eq. (\ref{ecuacion_estrella})
it follows that
$
(\Phi_1)_n \rightarrow \infty
$
as
$
n \rightarrow \infty,
$
unless
$
S_{11}(z)  =  0.
 $

We therefore seek solutions $z_1, z_2, \ldots, z_N$ of the equation
 \beq
 S_{11}(z_k) = 0,  \hspace{1cm}   k = 1, 2, \ldots, N,    \label{eival_eq}
 \eeq
such that  $|z_k|>1$. Then,
\[
(\Phi_1)_n(z_k)   = S_{21}(z_k) (\Psi_2)_n(z_k)  ,  \hspace{1cm}   k = 1, 2, \ldots, N.
\]
From this, it follows that $(\Phi_1)_n(z_k)$  decays   at $\pm \infty$:
\[
(\Phi_1)_n(z_k)  \rightarrow  0   \hspace{1cm}     \mbox{as}      \hspace{1cm}     n \rightarrow \pm\infty.
\]
It is then said that  $ (\Phi_1)_n(z_k)  $ is an {eigenfunction} ($k = 1, 2, \ldots, N$), with corresponding {eigenvalue} $z_k$.

In the case of $U_n(t=0)=U_0 \delta_{n,n_0}$,  the Jost function is
\beq
\Phi_n = Z^{n-1}(Z + M_0)\Phi_0, \hspace{0.5cm} \mbox{for} \hspace{0.5cm} n \geq 1,    \label{Phi_n}
\eeq
with $M_0$
\beq
M_0   =  \left(\!\begin{array}{ccc}  0 & U_0   \\ - U_0 & 0   \end{array}\!\right).       \label{eme_0}
\eeq
Furthermore, $\Psi_n = Z^n$ for $ n\geq1$ and  $\Phi_0$ is the identity
matrix: $\Phi_0 = I$. Hence, Eq. (\ref{Phi_n})  reads:
\[
\Phi_n = \Psi_n Z^{-1}(Z + M_0), \hspace{0.5cm} \mbox{for} \hspace{0.5cm} n \geq 1.
\]
Its comparison with Eq. (\ref{def_scat_matriz}) leads to a scattering
matrix
\[
S(z) = Z^{-1}(Z + M_0).
\]
We thus obtain that the transmission coefficient
\[
S_{11}(z)  =  1,
\]
which never vanishes. This means that  the one-site potential (i.e., a
single-site initial condition) does {\it not}
admit solitonic solutions, {\it independently} of the amplitude $U_0$ of
initial excitation. This theoretical result has also been confirmed by
numerical simulations for different values of $U_0$, always
leading to dispersion of the solution (not shown here).

\section{Threshold Conditions for the Non-Integrable DNLS Model
}

We now turn to the non-integrable DNLS lattice that has the
general form:
\begin{eqnarray}
i \dot{u}_n=-\epsilon \Delta_2 u_n - |u_n|^{2 \sigma} u_n ,
\label{deq1}
\end{eqnarray}
where $u_n$ is a complex field (corresponding to the envelope of the
electric field in optics, or the mean field wavefunction in optical
lattice wells in the BECs), $\Delta_2 u_n= (u_{n+1}+u_{n-1}-2u_n)$
is the discrete Laplacian, and $\epsilon$ is the ratio of the
tunneling strength to the nonlinearity strength.
Using a scaling invariance of the equation,
we can scale $\epsilon=1$, through rescaling
$t \rightarrow \epsilon t$ and $u_n \rightarrow
u_n/\sqrt{\epsilon}$.
Importantly for our
considerations, this model has an underlying Hamiltonian (which is
also a conserved quantity in the dynamics) of the form:
\begin{eqnarray}
H=\sum_n |u_{n+1}-u_n|^2 - \frac{1}{\sigma+1} |u_n|^{2 \sigma + 2}.
\label{deq2}
\end{eqnarray}
The only other known conserved quantity in the dynamics of
Eq. (\ref{deq1}) is the beam power (in Bose-Einstein condensates,
the normalized number of atoms) $P=\sum_n |u_n|^2$.

\begin{figure}[htbp]
\includegraphics[width=5cm,height=7cm,angle=0,clip]{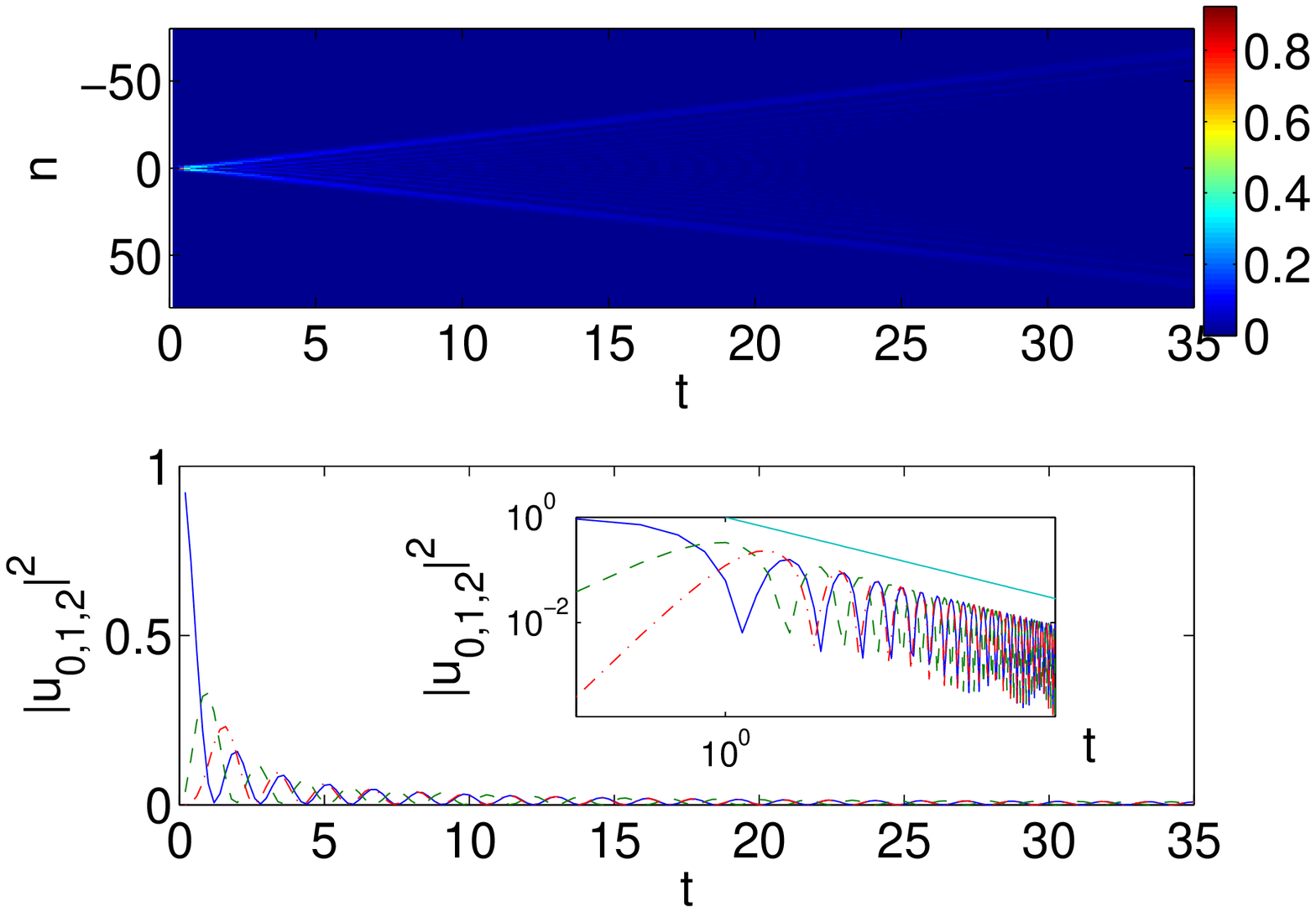}
\includegraphics[width=5cm,height=7cm,angle=0,clip]{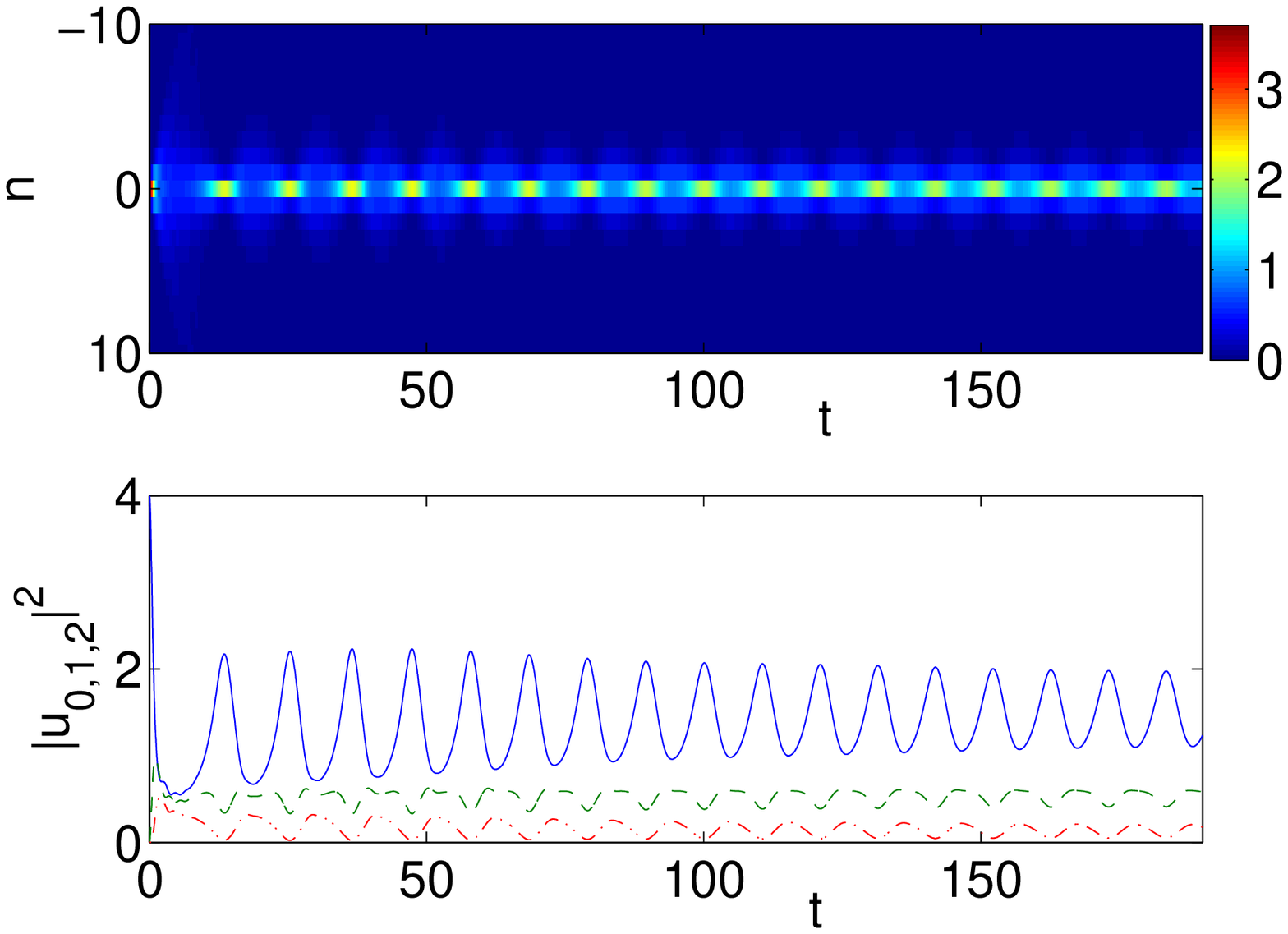}
\includegraphics[width=5cm,height=7cm,angle=0,clip]{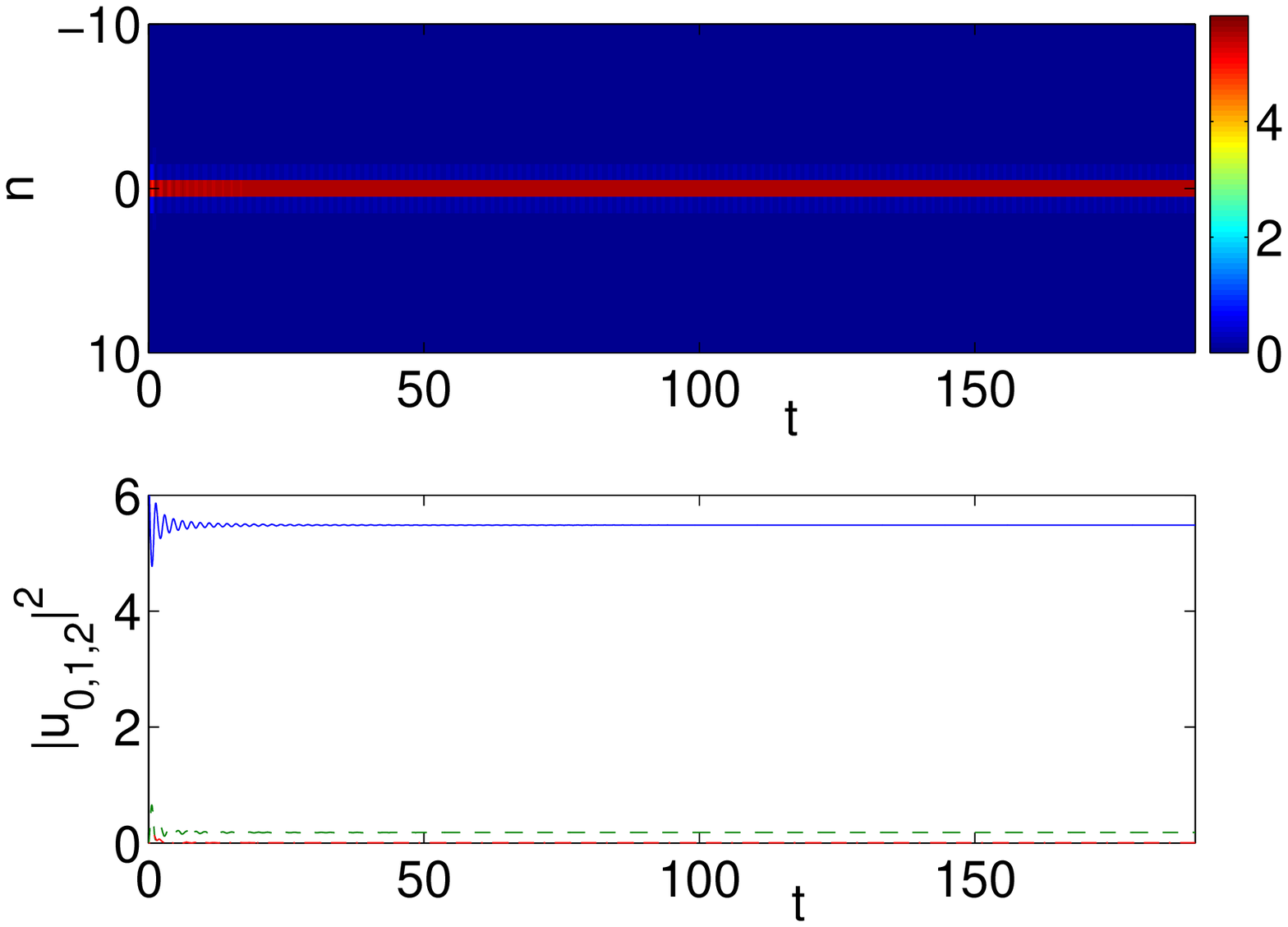}
\caption{Subcritical case (left), critical (middle), and
supercritical (right) initial single-site excitations on the
lattice. In all cases, the initial condition is $u_n=A
\delta_{n,0}$, with $A=1$ in the left, $2$ in the middle, and $2.5$
in the right panels. In each case the top panel shows the space-time
contour plot of the evolution. The bottom panel shows the dynamical
evolution of $|u_n|^2$ for sites $n=0,1,2$ (solid, dashed and
dash-dotted lines, respectively). For the leftmost case the inset
shows the same evolution in a log-log plot and a $t^{-1}$ decay for
comparison. Clearly, the damped oscillation of the field modulus has
an envelope of $t^{-1/2}$. } \label{dfig1}
\end{figure}

The specific problem that we examine here addresses the following
question. Consider a ``compactum'' of mass $u_n=A \delta_{n,0}$;
what is the critical value of $A$ that is necessary for this
single-site initial condition to excite a localized mode? That such
a threshold definitely exists is illustrated in Fig.~\ref{dfig1}.
The case of the leftmost panel is subcritical, leading to the
discrete dispersion of the initial datum. This follows the
well-known $t^{-1/2}$ amplitude decay which is implied by the
solution of the problem in the absence of the nonlinearity:
\begin{eqnarray}
u_n(t)=A i^n J_n(2 t) ,
 \label{deq3}
\end{eqnarray}
where $J_n$ is a Bessel function of order $n$. Notice
that this can also be shown to be consistent with the findings
of \cite{ournon}.
On the other hand, the rightmost panel shows a
nonlinearity-dominated regime with the rapid formation of a solitary
wave strongly localized around $n=0$. In the intermediate case of
the middle panel, the system exhibits a long oscillatory behavior
reminiscent of a separatrix between the basins of attraction of the
two different regimes.

\begin{figure}[htbp]
\includegraphics[width=7cm,height=7cm,angle=0,clip]{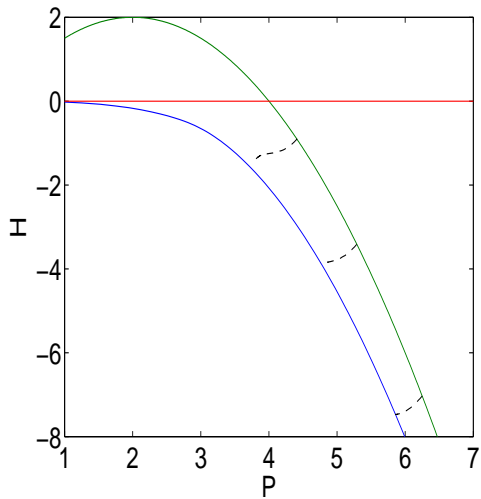}
\caption{The blue solid line shows the energy $H$ versus the power
$P$ of the discrete solitary wave solutions. Above it, the green
solid line shows the energy versus power of the initial condition,
obtained from Eq. (\ref{deq4}) and $P_{ss}=A^2$. The horizontal line
denotes $H=0$ and its intersection with the initial condition curve
defines the single-site, initial amplitude $A^{\ast}$. The
dynamical evolution of three different supercritical initial states
with $A=2.1$, $A=2.3$, and $A=2.5$ is shown by dashed (black) lines;
see also discussion in the text.} \label{dfig2}
\end{figure}

We argue herein that the essence of this separatrix lies in the
examination of the stationary (localized) states of the model. Such
standing wave solutions of the form $u_n=\exp(i \Lambda t) v_n$,
which are exponentially decaying for $v_n$ as a function of $n$, can
be found for arbitrary frequency $\Lambda$ (and arbitrary power $P$
in one spatial dimension). This is a well-known result in one
dimension; see e.g., Ref.~\cite{johanson}. On the other hand, the
single-site initial condition discussed above has an energy of:
\begin{eqnarray}
H_{ss} = 2 A^2 - \frac{1}{\sigma+1} A^{2 \sigma + 2},
\label{deq4}
\end{eqnarray}
where the subscript denotes single site. Figure~\ref{dfig2}
summarizes succinctly the power dependence of the energy for these
two cases for $\sigma=1$. Both the energy of the stationary
solutions as a function of their power and the single-site energy as
a function of single-site power ($P_{ss} = A^2$),
are shown. Note that the two curves $H_{ss} (P_{ss})$ and $H(P)$ do
not intersect (except at the trivial point $H=P=0$) since for
$\epsilon \neq 0$ single-site states are not stationary ones.

The examination of Fig.~\ref{dfig2} provides information on the existence
of a sufficient condition for the formation of a localized mode and on the
dynamics of the single-site initial condition. Figure~\ref{dfig2}
shows that localized solutions exist for arbitrarily small values of
the input power $P$, and that the energy of the localized states is
negative.
This implies that the crucial quantity
to determine the fate of the process is the energy $H$
and not the power $P$. The role of the power will become more
evident in the two-dimensional setting mentioned below. Moreover, if
the system starts at a given point on the curve defined by $H_{ss}=2
P_{ss} - P_{ss}^{\sigma+1}/(\sigma+1)$, due to conservation
of total $H$ and total $P$, it can only end up in a stationary state
in the quadrant $H < H_{ss}$ and $P < P_{ss}$. That is to say,
some of the initial energy and power are typically ``shed off''
in the form of radiation (i.e., converted to other degrees of
freedom which is the only way that ``effective dissipation'' can
arise in a purely Hamiltonian system), so that the initial condition
can ``relax'' to the pertinent final configuration.
As mentioned above, a localized solution with the same power as that of the
initial condition exists for arbitrary $A$. {\it However},
emergence of a localized mode occurs only
for those initial conditions whose core energy (i.e., the
energy of a region around the initially excited state) is negative,
 after the profile is ``reshaped'' by
radiating away both energy and power.

Therefore, if $H_{ss}<0$, then the compactum of initial data will
always yield a localized excitation: this inequality provides the
sufficient condition for the excitation of solitary waves.
The condition on the energy, in turn, provides a condition on the single-site
amplitude that leads to the formation of solitary waves, namely,
solitary waves always form if $A > A^{\ast}$ with
\begin{eqnarray}
A^{\ast} =\left[2 \left(\sigma+1\right)
\right]^{\frac{1}{2 \sigma}}. \label{deq5}
\end{eqnarray}
For the case of $\sigma=1$ considered in Figs.~\ref{dfig1}
and \ref{dfig2}, this amplitude value is $A^{\ast} =2$ in agreement
with our numerical observations of Fig.~\ref{dfig1}. Whether an
initial state with $H_{ss}>0$ yields a localized state depends on
the explicit system dynamics, corroborating the observation that the
previous energetic condition is a sufficient, but not necessary, condition.

One should make a few important observations here.
Firstly, we note the stark contrast of the non-integrable
discrete model and both of its integrable (continuum and
discrete) counterparts. In the (singular) continuum limit, it is possible
to excite a single soliton or a multi-soliton depending on
the barrier height and width. On the other hand, in the integrable
discretization one-site excitation {\it never} leads to solitary wave
formation, contrary to what is the case here where either none or
one solitary wave may arise, depending on the amplitude of the initial
one-site excitation.

Restricting our consideration to the non-integrable model,
we observe that even
though a localized solution with the same power as that of the
initial condition exists for arbitrary $A$, formation of a localized
mode will always
occur only for $A>A^{\ast}$, as determined by the previous
energetic criterion.  Secondly, the answer that the formation always
occurs for $A>A^{\ast}$ generates two interconnected
questions: what is the threshold for the formation of the localized
mode, and given an
initial $A>A^{\ast}$, which one among the mono-parametric
family of solutions will the dynamics of the model select as the end
state of the system ? [It should be noted in connection to the latter
question that single-site initial conditions have {\it always} been
found in our numerical simulations to give rise to {\it at most} a
single-site-centered solution i.e., multipulses cannot be produced
by this process.]
Some examples of this dynamical process are
illustrated in Fig. \ref{dfig2}, where the energy and power of a few
sites (typically 20-40) around the originally excited one are
measured as a function of time and are parametrically plotted in the
$H-P$ plane. As it should, the relevant curve starts from the
$H_{ss}-P_{ss}$ curve, and asymptotically approaches, as a result of
the dynamical evolution, the $H-P$ curve of the stationary states of
the system. However, the relaxation process happens {\it neither at
fixed energy, nor at fixed power}. Instead, it proceeds through a
more complex, dynamically selected pathway of loss of both $H$ and
$P$ to relax eventually to one of the relevant stationary states.
This is shown for three different values of super-critical amplitude
in Fig. \ref{dfig2} ($A=2.1$, $2.3$ and $2.5$). We have noticed
(numerically) that the loss of energy and power, at least in the
initial stages of the evolution happens at roughly the same rate,
resulting in $dH/dP \approx {\rm const}$. However, the later stages
of relaxation no longer preserve this constant slope.
It is worthwhile to highlight here that the fact that the
process of formation of a localized mode is neither equienergetic,
nor does it occur at fixed power is something that has been
previously observed in the continuum version of the system
in \cite{akm2} (cf. with the discussion in p. 6095 therein).

The dynamics of the system in the $H-P$ space can be qualitatively
understood by considering the frequencies $\Lambda$ associated with
the instantaneous $H$, $P$ values along the system trajectories. At
each instantaneous energy $H$ a frequency $\Lambda_H$ may be defined
as the frequency of a single-site breather stationary state with
energy $H$. Such a frequency is unique as the stationary-state
energy is a monotonically decreasing function of the
frequency~\cite{johanson}. Similarly, a frequency $\Lambda_P$ may be
uniquely associated with the instantaneous power $P$, i.e.,
$\Lambda_P$ is the frequency of a stationary state with power $P$.
Note, however, that the stationary-state power is a monotonically
increasing function of $\Lambda$. At the final stationary state
where the system relaxes $\Lambda_H = \Lambda_P = - dH/dP$. Hence,
the system trajectories are such that $\Lambda_H$ increases
(consequently the energy decreases) and $\Lambda_P$ decreases (the
power decreases). The final stationary state is reached when the two
frequencies become equal, the point where they meet depending on
their rate of change along the trajectory, i.e., on their
corresponding ``speeds'' along the trajectory.
Nevertheless, the precise mechanism of selection of the particular
end state (i.e., of the particular ``equilibrium $\Lambda$'')
that a given initial state will result in remains a formidable
outstanding question that would be especially interesting
to address in the future.



\begin{figure}[htbp]
\includegraphics[width=5cm,height=7cm,angle=0,clip]{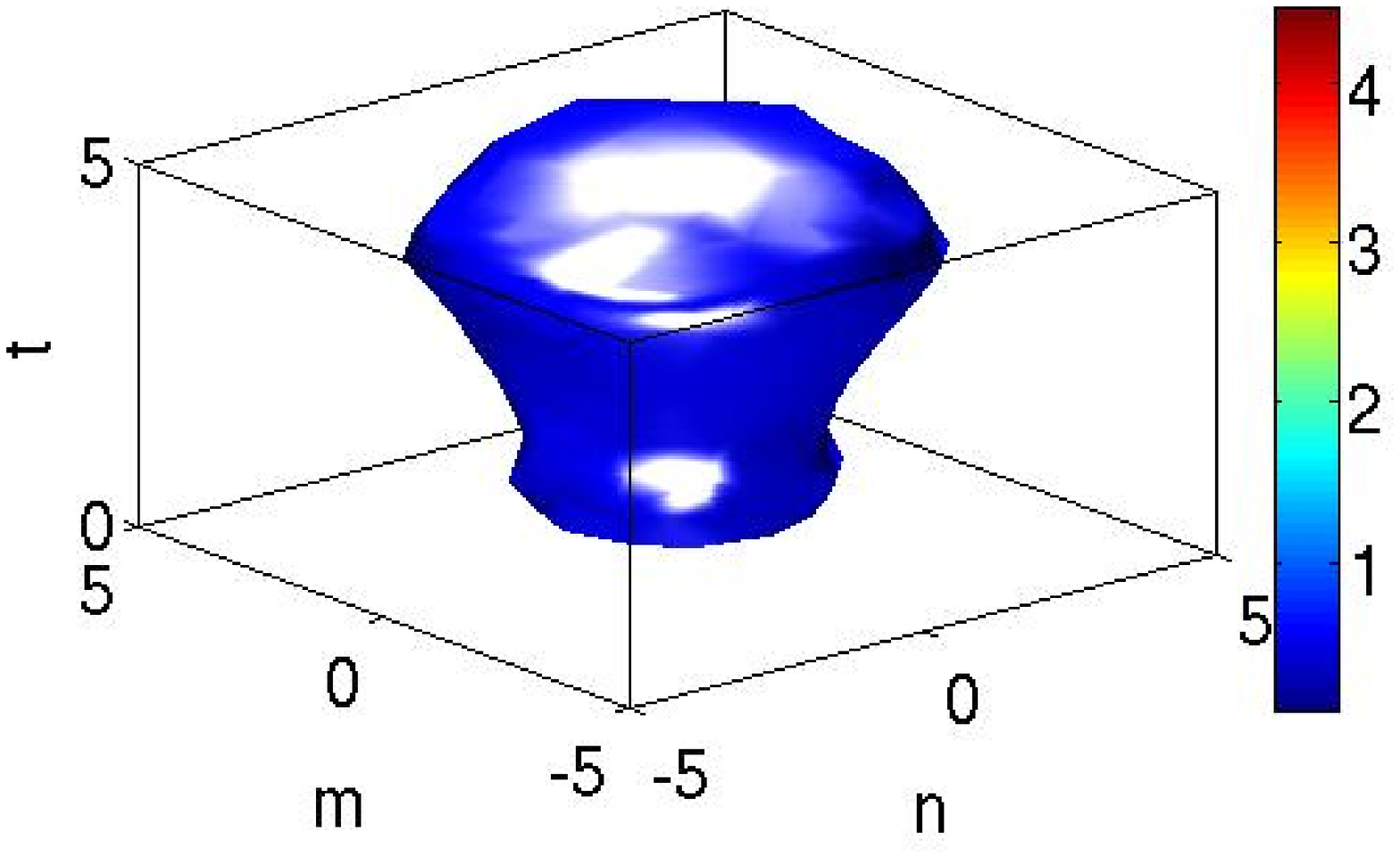}
\includegraphics[width=5cm,height=7cm,angle=0,clip]{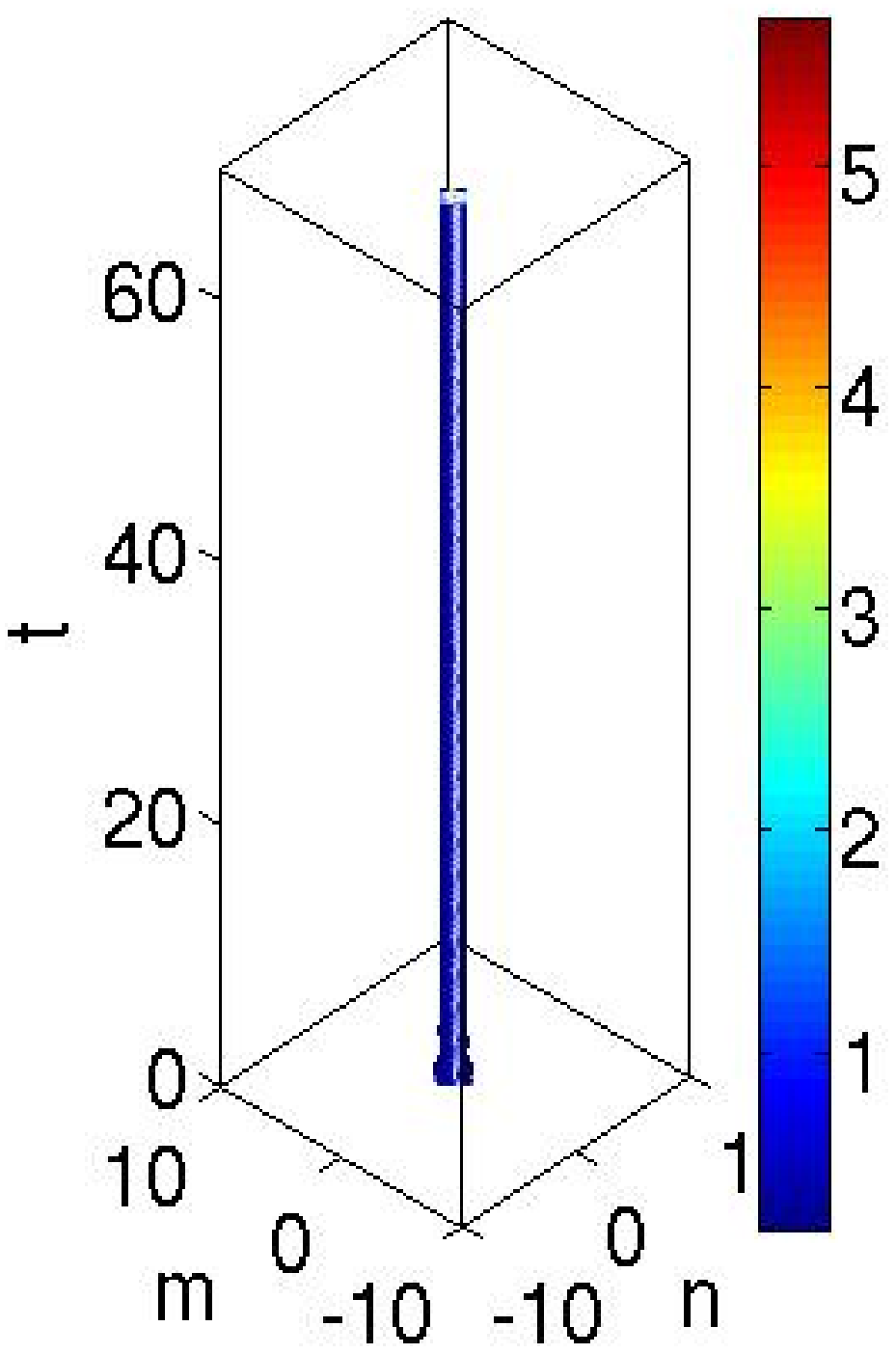}
\includegraphics[width=5cm,height=7cm,angle=0,clip]{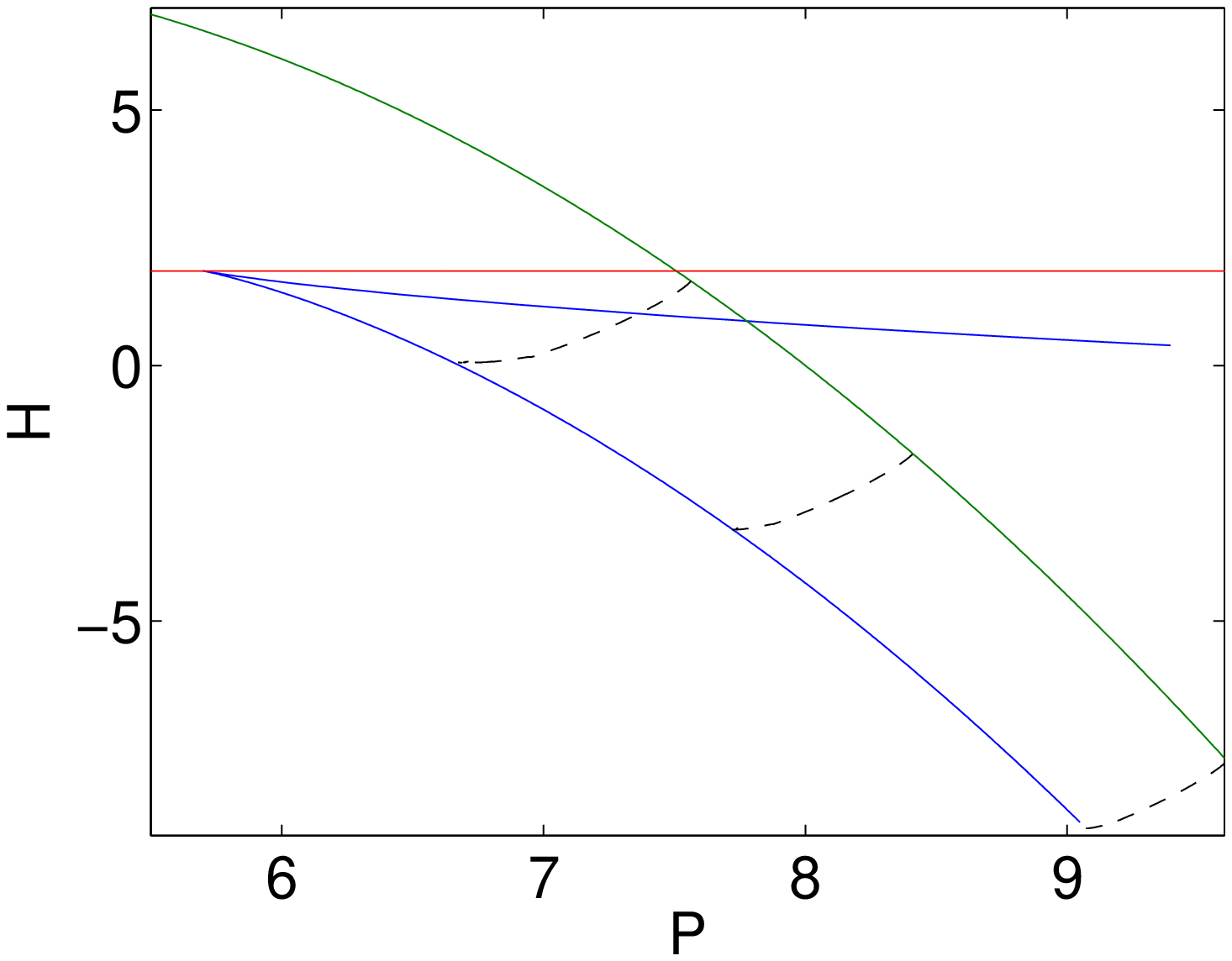}
\caption{The left panel shows the spatio-temporal evolution of
$|u_{n,m}(t)|^2$ for the sub-critical case with amplitude $A=2.65$.
The middle panel shows the weakly super-critical case of $A=2.75$.
Finally the right panel is analogous to Fig. \ref{dfig2}, showing
the $H-P$ diagram for the solution branch (wedge-like blue solid
line), the $H_{ss}-P_{ss}$ graph of the initial conditions (green
solid line) and the trajectories of three super-critical cases in
$H-P$ space for $A=2.75$, $2.9$ and $3.1$ (dashed black lines).
The red horizontal line represents the maximal energy for which solutions are
found to exist, namely $H_0 \approx 1.85$.}
\label{dfig3}
\end{figure}

We now turn to the two-dimensional variant of the above
one-dimensional non-integrable lattice.
Equation~(\ref{deq1}) remains the same, but for the two-dimensional
field $u_{n,m}$, and the discrete Laplacian becomes the five-point
stencil $\Delta_2 u_{n,m} = (u_{n+1,m}+u_{n-1,m}+
u_{n,m+1}+u_{n,m-1}-4 u_{n,m})$. For the initial condition
$u_{n,m}=A \delta_{n,0} \delta_{m,0}$, the $H_{ss}-P_{ss}$ curve is
given by:
\begin{eqnarray}
H_{ss}=4 P_{ss}-P_{ss}^{\sigma+1}/(\sigma+1). \label{deq7}
\end{eqnarray}
In $d$-dimensions the first term would be $2 d  P_{ss}$. We
once again find the branch of standing waves of the equation. While
this branch of solutions is also well-known~\cite{kevrek}, there are
some important differences with the one-dimensional case. Firstly, a
stable and an unstable branch of solutions exists (per the
well-known Vakhitov-Kolokolov criterion); in the wedge-like curve
indicating the standing waves only the lower energy branch is
stable. Furthermore, the maximal energy of the solutions is no
longer $H=0$, but finite and positive (in fact, for
$\sigma=1$, it is $H_0 \approx 1.85$). Finally, solutions
no longer exist for arbitrarily low powers, but they may only exist
above a certain power (often referred to as the excitation threshold
\cite{excit1,excit2,excit3}). We can now appreciate the impact of
these additional features in the right panel of Fig.~\ref{dfig3}.
Comparing $H_{ss}$ with $H_0$ we find two solutions: one with $A
\approx 0.7$ and one with $A \approx 2.73$. However, for the lower
one there are no standing wave excitations with the corresponding
power (this reveals the role of the power in the higher dimensional
problem). Hence, the relevant amplitude that determines the
sufficient condition in the two-dimensional
case is the latter. Indeed, the case with $A=2.65$ shown in the left
panel of Fig.~\ref{dfig3} shows spatio-temporal diffraction, while
that of $A=2.75$ in the middle panel illustrates robust
localization, indicating the super-critical nature of the latter
case. Notice that the pathway of the dynamics is presented in the
right panel for three super-critical values of $A=2.75$, $A=2.9$ and
$A=3.1$. Once again, the dynamics commences on the $H_{ss}-P_{ss}$
curve, as it should, eventually relaxing on the {\it stable}
standing wave curve. As before, the initial dynamics follows a
roughly constant $dH/dP$, but the relaxation becomes more complex at
later phases of the evolution.

\section{Conclusions and Future Challenges}

The above study has examined the presence of a sharp crossover
between the linear and nonlinear dynamics of a prototypical
dynamical lattice model such as the discrete nonlinear
Schr{\"o}dinger equation. This crossover has already been observed
in media with the focusing nonlinearity~\cite{mora} (as considered
here). It has also been observed very recently in media with the
defocusing nonlinearity~\cite{rosberg}. The latter can be
transformed into the former under the so-called staggering
transformation $u_n=(-1)^n w_n$, where $w_n$ is the field in the
defocusing case. As a result, the solitary waves of the defocusing
problem discussed in \cite{rosberg} will be ``staggered'' (i.e., of
alternating phase between neighboring sites), yet the phenomenology
discussed above will persist. The crossover was quantified on the
basis of an energetic comparison of the initial-state energy with
the branch of corresponding stable localized solutions ``available'' in the
model. A sufficient, but not necessary, condition for the excitation
of a localized mode based on the initial-state, single-site
amplitude was discussed, and it was successfully tested in numerical
simulations.
Similar findings were obtained in the two-dimensional analog of the problem:
a crossover behaviour dictated by the energy was found, but
the crossover was also affected by the power and its excitation thresholds.
Furthermore, these results were contrasted with the case
of the continuum version of the model, where depending on the
strength of the excitation, also multi-solitons can be obtained
and with the integrable discrete Ablowitz-Ladik model where
no single-site
excitation can produce a solitary wave,
independently of the excitation amplitude.

However, a number of interesting questions emerge from these
findings that are pertinent to future studies. Perhaps the foremost
among them concerns how the dynamics ``selects'' among the available
steady-state excitations with energy and power below that of the
initial condition the one to which the dynamical evolution leads.
According to our results, this evolution is neither
equienergetic, nor power-preserving (see also \cite{akm2}),
hence it would be extremely
interesting to identify the leading physical principle which
dictates it. Another question generalizing the more principal one
asked herein concerns the excitation of multiple sites (possibly at
the same amplitude), starting with two, and inquiring the nature of
the resulting state of the system. On a related note, perhaps this
last question is most suitable to be addressed first in the
context of the integrable model where the formulation presented herein
can be generalized and yield definitive answers for the potential
excitation of solitons.
Studies along these directions are
currently in progress and will be reported in future publications.


\section{Acknowledgements}

YD gratefully acknowledges discussions with L. Isella.
PGK gratefully acknowledges support from grants NSF-DMS-0505663,
NSF-DMS-0619492 and NSF-CAREER.


\begin{thebibliography}{99}

\bibitem{reviews} S. Aubry,  Physica D {\bf 103},
201, (1997); S. Flach and C.R. Willis,  Phys. Rep.
{\bf 295}, 181 (1998);
D. Hennig and G. Tsironis,
Phys. Rep. {\bf 307}, 333 (1999);

\bibitem{optics} D. N. Christodoulides, F. Lederer and Y. Silberberg,
Nature \textbf{424}, 817 (2003); Yu. S. Kivshar and G. P. Agrawal,
\textit{Optical Solitons: From Fibers to Photonic Crystals},
Academic Press (San Diego,
2003).


\bibitem{bec_reviews}
V.V. Konotop and V.A. Brazhnyi,
Mod. Phys. Lett. B {\bf 18}, 627 (2004); O. Morsch and M. Oberthaler,
Rev. Mod. Phys. {\bf 78}, 179 (2006);
P.G. Kevrekidis and D.J. Frantzeskakis,
Mod. Phys. Lett. B {\bf 18}, 173 (2004).

\bibitem{sievers} M. Sato, B. E. Hubbard, and A. J. Sievers,
Rev. Mod. Phys. {\bf 78}, 137 (2006).

\bibitem{peyrard} M. Peyrard,
Nonlinearity {\bf 17}, R1 (2004).

\bibitem{dnls}  P.G. Kevrekidis, K.O. Rasmussen, and A.R.
Bishop,  Int. J. Mod. Phys. B {\bf 15}, 2833 (2001).

\bibitem{mora} H.S. Eisenberg, Y. Silberberg, R. Morandotti,
A.R. Boyd and J.S. Aitchison, Phys. Rev. Lett. {\bf 81}, 3383 (1998).

\bibitem{rosberg} M. Matuszewski, C.R. Rosberg, D.N. Neshev,
A.A. Sukhorukov, A. Mitchell, M. Trippenbach, M.W. Austin,
W. Krolikowski and Yu.S. Kivshar,
Opt. Express {\bf 14}, 254 (2006).

\bibitem{sulem} C. Sulem and P.L. Sulem,
\newblock  {\it The Nonlinear Schr{\"o}dinger Equation},
Springer-Verlag (New York, 1999).


\bibitem{ZS} V.E. Zakharov and A.B. Shabat,
Soviet Physics-JETP {\bf 34}, 62 (1972).

\bibitem{neshev} J.W. Fleischer, T. Carmon, M. Segev, N.K. Efremidis
and D.N. Christodoulides,
Phys. Rev. Lett. {\bf 90}, 023902 (2003);
J.W. Fleischer, M. Segev, N.K. Efremidis and D.N. Christodoulides,
Nature {\bf 422}, 147 (2003).


\bibitem{fleischer} H. Martin, E.D. Eugenieva,
Z. Chen and D.N. Christodoulides,
Phys. Rev. Lett. {\bf 92}, 123902 (2004).


\bibitem{neshev2} D.N. Neshev, T.J. Alexander, E.A. Ostrovskaya,
Yu.S. Kivshar, H. Martin, I. Makasyuk and Z. Chen,
Phys. Rev. Lett. {\bf 92}, 123903 (2004).

\bibitem{fleischer2} J.W. Fleischer, G. Bartal, O. Cohen, O. Manela,
M. Segev, J. Hudock and D.N. Christodoulides,
Phys. Rev. Lett. {\bf 92}, 123904 (2004).

\bibitem{moti} J.W. Fleischer, G. Bartal, O. Cohen, T. Schwartz, O. Manela,
B. Freeman, M. Segev, H. Buljan, and N.K. Efremidis,
Optics Express {\bf 13}, 1780 (2005).

\bibitem{al1} M.J. Ablowitz and J.F. Ladik,
J. Math. Phys. {\bf 16}, 598 (1975).

\bibitem{al2} M.J. Ablowitz and J.F. Ladik,
J. Math. Phys. {\bf 17}, 1011 (1976).

\bibitem{trub} M.J. Ablowitz, B. Prinari and A.D. Trubatch,
{\it Discrete and Continuous Nonlinear Schr{\"o}dinger Systems},
Cambridge University Press (Cambridge, 2004).

\bibitem{akm1} N.N. Akhmediev and A. Ankiewicz, {\it Solitons:
Nonlinear pulses and beams}, Chapman \& Hall (London 1997).


\bibitem{akm2} N.N. Akhmediev, A. Ankiewicz, and R. Grimshaw,
Phys. Rev. E {\bf 59}, 6088 (1999).

\bibitem{rumpf} B. Rumpf,
Phys. Rev. E {\bf 70}, 016609 (2004).


\bibitem{klaus2}   M. Klaus and J. K. Shaw,
Phys. Rev. E {\bf 65}, 036607 (2002).

\bibitem{ournon} A. Stefanov and P.G. Kevrekidis,
Nonlinearity {\bf 18}, 1841 (2005).

\bibitem{johanson} M. Johansson and S. Aubry,
Phys. Rev. E {\bf 61}, 5864 (2000).

\bibitem{kevrek} P. G. Kevrekidis, K. {\O}. Rasmussen, and A. R. Bishop,
Phys. Rev. E {\bf 61}, 2006-2009 (2000).

\bibitem{excit1} S. Flach, K. Kladko and R.S. MacKay,
Phys. Rev. Lett. {\bf 78}, 1207 (1997).

\bibitem{excit2} M.I. Weinstein, Nonlinearity {\bf 12},
673 (1999).

\bibitem{excit3} P.G. Kevrekidis, K.{\O}. Rasmussen
and A.R. Bishop,
Phys. Rev. E {\bf 61}, 4652 (2000).






\end{thebibliography}
\end{document}